\documentclass[twocolumn,showpacs,preprintnumbers,amssymb]{revtex4}

\usepackage{graphics}
\usepackage{epsfig} 
\usepackage{hyperref}
\usepackage{graphics}
\usepackage{color}      
\usepackage{graphicx}
\usepackage{graphicx}
\usepackage{dcolumn}

\begin{document}

\title{ Exact time-dependent analytical  solutions for  entropy  production rate  for a system that operates in a heat bath where its temperature varies linearly in space}
\author{Mesfin Asfaw  Taye}
\affiliation {West Los Angeles College, Science Division \\9000  Overland Ave, Culver City, CA 90230, USA}

\email{tayem@wlac.edu}

\begin{abstract}

The nonequilibrium thermodynamics feature of 
 a Brownian motor is investigated by obtaining exact time-dependent solutions. This in turn enables  us to investigate not only
the long time property (steady-state) but also the short time
 the behavior of the system.  The general expressions for the free energy, entropy production ${\dot e}_{p}(t)$ as well as entropy extraction ${\dot h}_{d}(t)$ rates are derived for a system that is genuinely driven out of equilibrium by time-independent force as well as by spatially varying thermal background.   We show that for a system that operates between hot and cold reservoirs,  most of the thermodynamics quantities approach a non-equilibrium steady state in the long time limit. The change in free energy becomes minimal at a steady state. However for a system that operates in a heat bath where its temperature varies linearly in space, the entropy production and extraction rates approach a non-equilibrium steady state while the change in free energy varies linearly in space. This reveals that unlike 
systems at equilibrium,  when 
systems are driven out of equilibrium,   their free energy may not be minimized.  
The thermodynamic properties of a system that operates between the hot and cold baths are further compared and contrasted with a system that operates in a heat bath where its temperature varies linearly in space along with the reaction coordinate.  We show that the entropy, entropy production, and extraction rates are considerably larger for linearly varying  temperature case than a system that  operates between the hot and cold baths revealing such systems are inherently irreversible. For both cases, in the presence of load or when a distinct temperature difference is retained,  the entropy $S(t)$ monotonously increases with time and saturates to a constant value as $t$ further steps up.  The entropy production rate ${\dot e}_{p}$ decreases in time and at steady state, ${\dot e}_{p}={\dot h}_{d}>0$ which agrees with the results shown in the works \cite{mu17,muu17}. Moreover, the velocity, as well as the efficiency of the system that operates between the hot and cold baths, are also collated and contrasted with a system that operates in a heat bath where its temperature varies linearly in space along with the reaction coordinate. A system that operates between the hot and cold baths has significantly lower velocity but a higher efficiency in comparison with a linearly varying  temperature case. 

\end{abstract}
\pacs{Valid PACS appear here}
\maketitle


 \section{Introduction}

Thermodynamics is one of the 
most studied disciplines since its applications 
encompass a variety of topics in science and engineering.  It can be further subdivided into equilibrium and non-equilibrium disciplines. 
Equilibrium thermodynamics is well studied 
but has limited applications
since most systems in nature are far from equilibrium. In this case, 
its macroscopic properties can be further verified from a microscopic point of view via equilibrium statistical mechanics.  In contrast,   nonequilibrium thermodynamics deals with inhomogeneous systems where the system thermodynamic quantities rely on the reaction rates in a  complicated manner. As a result, getting a universal exact result was unattainable. However,  in  the last few decades,  several studies have been conducted to explore the nonequilibrium thermodynamic feature of systems that are out of equilibrium \cite{mu1,mu2,mu3,mu4, mu5,mu6,mu7,mu8,mu9,mu10,mu11,mu12,ta1,mu13,mu14,mu15,mu16}. Some notable works in this regard include,   the analytically solvable models depicted in the works \cite{mu17,muu17} and the study of thermodynamic features for systems that operate in the quantum realm \cite{mu25,mu26,mu27}.  Furthermore,  for systems that are genuinely driven out of equilibrium,  the thermodynamic relations were derived for a Brownian particle that walks in an overdamped medium \cite{muuu17} and underdamped medium \cite{muuu177}.  The method of calculating entropy production and extraction rates at ensemble level by first analyzing the thermodynamic relation at trajectory level was introduced in the work \cite{mu6}.  Alternatively,  many thermodynamic relations were reconfirmed under 
time reversal operation \cite{mar2,mar1}. Such studies help to 
comprehend the thermodynamic properties of biological systems such as intracellular transport of kinesin or dynein inside the cell  \cite{mu28,mu29,mu30}.

Since real systems operate in a finite time,  solving the model system exactly as a function of time is fundamental to grasp the thermodynamic features of the systems beyond a linear response and steady-state regimes. In  this work by obtaining exact time-dependent solutions, 
we  investigate not only
the long time property (steady-state) but also the short time
the behavior of the system.  The general expressions for free energy, entropy production as well as entropy production rates are derived for a system that is genuinely driven out of equilibrium by time-independent force as well as by spatially varying thermal background. By solving the model as a function of time,  the dependence of these thermodynamic quantities as a function of time is explored.  For a system that operates between hot and cold reservoirs,  most of these thermodynamics quantities approach a non-equilibrium steady state in the long time limit. The change in free energy becomes minimal at a steady state. However for a system that operates in a heat bath where its temperature decreases linearly, the entropy production and extraction rates approach a non-equilibrium steady state while the change in  the free energy decreases linearly. This reveals  that unlike 
systems at  equilibrium,  when 
systems are driven out of equilibrium,   their free energy may keep decreasing as time evolves.  In the absence of load and isothermal cases, we show that the non-equilibrium state relaxes to equilibrium.

In this work,  we also consider a   simple model where the single-particle walks in one-dimensional discrete ratchet potential with a load. The ratchet potential is also coupled with a heat bath. The thermodynamic properties of a system that operates between the hot and cold baths are compared and contrasted with a system that operates in a heat bath where its temperature linearly decreases along with the reaction coordinate.  We show that the entropy ${ S}(t)$, the entropy production ${\dot e}_{p}(t)$ and extraction rates ${\dot h}_{d}(t)$   are considerably larger for linearly decreasing temperature case than a Brownian particle that operates between the hot and cold baths revealing such systems are inherently irreversible. For both cases, in the presence of load or when a distinct temperature difference is retained,  the entropy $S(t)$ monotonously increases with time and saturates to a constant value as $t$ further steps up.  The entropy production rate ${\dot e}_{p}$ decreases in time and at steady state, ${\dot e}_{p}={\dot h}_{d}>0$ which agrees with the results shown in the works \cite{mu17,muu17}. On the contrary, for an isothermal case and in the absence of load, ${\dot e}_{p} = {\dot h}_{d}=0$ in a long time limit which is a reasonable argument as any system which is in contact with a uniform temperature should obey the detail balance condition.

Because closed-form expressions for 
${\dot S}(t)$, ${\dot e}_{p}(t)$  and ${\dot h}_{d}(t)$ as a function of  $t$ are obtained, 
the analytic expressions for  the change in entropy production $\Delta e_p(t)$,  heat dissipation $\Delta h_d(t)$ and  total entropy $\Delta S(t)$ can be found. We show that for a system that operates between hot and cold reservoirs,  $\Delta h_d(t)$  and $\Delta e_p(t)$ approach a non-equilibrium steady state in the long time limit.
 However, for a system that operates in heat baths where its temperature decreases linearly, $\Delta h_d(t)$ and $\Delta e_p(t)$ increase linearly as time progresses. In the absence of a load, potential barrier, and for isothermal case, for both cases,  $\Delta S=\ln [3]$ which reconfirms the well-known relation for a system under infinitesimal process.  In other words,  since the system has three accessible states (three lattices) $\Omega=3$,  at equilibrium $S=\ln(\Omega)$. At Equilibrium one also  finds, ${\dot e}_{p}(t)=\ln [3]$  and ${\dot h}_{d}(t)=0$. Moreover, the change in free energy   ${\Delta F}$  decreases in time and saturates to a constant but minimal value for the system that operates between the hot and cold baths.  On the contrary, for the system that operates in a linearly decreasing temperature case,  the free energy decreases linearly.

The velocity, as well as the efficiency of  the system that operates between the hot and cold baths, are also compared and contrasted with a system that operates in a heat bath where its temperature linearly decreases along with the reaction coordinate. A system that operates between the hot and cold baths has significantly lower velocity but a higher efficiency in comparison with a linearly decreasing temperature  case.  For a  linearly decreasing  temperature case,  we show that the efficiency of such a Brownian heat engine is far less than   Carnot's efficiency even at the quasistatic limit.  At quasistatic limit, the efficiency  of the  heat engine approaches  the efficiency of endoreversible engine  $\eta=1-\sqrt{{T_{c}/T_{h}}}$  \cite{mg1}. Moreover, the dependence of the current, as well as the efficiency on the model parameters, is explored analytically.

The rest of the paper is organized as follows: in Section II, we present the model and derive  
the expression for various thermodynamic relations for a Brownian particle walks in one-dimensional discrete ratchet potential with a load.   In Section III, the role of time on entropy and free energy is explored. In section IV, the dependence of efficiency and velocity on model parameters is explored.  Section V deals with the summary and conclusion.

\section{ The model and  derivation of free energy}

In this section, we derive the general expression for free energy, entropy production as well as entropy production rates for a system that is  driven out of equilibrium by time-independent force as well as spatially varying thermal background. By solving the model as a function of time, we explore the dependence of these thermodynamic quantities as a function of time.  Let us now  consider a  Brownian particle that moves in a  discrete  lattice   where its   dynamics  is  governed by   the master equation \cite{mu17} 
\begin{equation} 
{dP_{n} \over dt}=\sum_{n\neq n'}\left(P_{nn'}p_{n'}-P_{n'n}p_{n}\right),~~n,n'=1,2,3.
\end{equation} 
Here  
$P_{n'n}$  is the transition probability rate at which the system, originally in state $n$, makes a transition to state $n'$. $P_{n'n}$ is given by the Metropolis rule \cite{mu17}. Next, the relation for the entropy  production rate as well as the free energy   will be explored   as a  function of time   by considering a Brownian particle that moves   along  one dimensional  discrete ratchet potential.
 \begin{figure}[ht]
\centering
{
    \includegraphics[width=6cm]{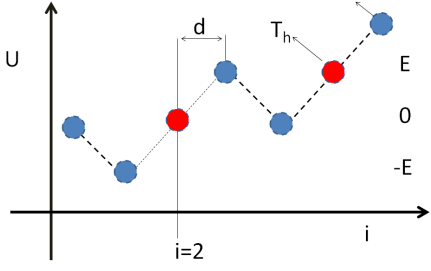}}
\caption{(Color online)  Schematic diagram for a Brownian particle  walking in a discrete ratchet potential with load. 
Sites with red  circles are coupled to the hot reservoir ($T_h$)
while sites with blue circles are coupled to the cold reservoir
($T_c$). Site 1 is labeled explicitly and $d$ is the lattice spacing. } 
\end{figure}

{\it Case1:  Brownian particle  operating between   hot and cold reservoirs.\textemdash}
Before  considering a linearly  decreasing temperature  profile, for clarity   let us  first rederive the entropy production  rate   for a Brownian particle   that moves in one dimensional  discrete ratchet potential  $U_i$  \cite{mu17}
\begin{equation} 
U_i = E[i(mod)3 - 1]+ifd.
\end{equation} 
The ratchet potential is  coupled  with the temperature
\begin{equation} 
  T_i=\{\begin{array}{cl}
   T_{h},&if~ E[i(mod)3 - 1]=0;\\
   T_{c},&otherwise;\end{array}
   \end{equation}
as shown in Fig. 1. The potential   $E>0$, $f$ denotes the load and     $i$ is an integer
that runs from $-\infty$ to $\infty$.  $T_{h}$ and $T_{c}$ designate the temperature for the hot and cold reservoirs, respectively.  Moreover, $d$ denotes the  lattice   spacing $d$ and in one cycle, the particle walks a net displacement of three  lattice sites 
as shown in Fig. 1.  The jump probability   from site $i$ to $i+1$ is given by 
$\Gamma e^{-\Delta E/k_{B}T_{i}}$
 where $\Delta E = U_{i+1} - U_i$ and $\Gamma$
is the probability attempting a jump per unit time.  $k_{B}$ designates the Boltzmann constant and  hereafter $k_{B}$, $\Gamma$  and $d$ are  considered  to be a unity. Obeying the metropolis algorithm
when  $\Delta E \le 0$,  
the jump  definitely takes place while $\Delta E > 0$  the jump
takes place with probability $\exp(-\Delta E/T_{i})$ \cite{mu13}.
Substituting the  probability rates (see Eq. (6)) into Eq. (1) yields 
\begin{equation} 
{d \vec{p}  \over dt}= {\bold P}\vec{p}
\end{equation} 
where $\vec{p}=(p_{1},p_{2},p_{3})^T$. 
${\bold P}$ is a 3 by 3 matrix which
is given by
\begin{equation} 
{\bold P}= \left( \begin{array}{ccc}
{-\mu a^2-\mu^2\over 2a} & {1\over 2} & {1\over 2} \\
{\mu a \over 2} & {-1-\nu b\over 2} & {1\over 2}\\
{\mu^2 \over 2a} & {\nu b\over 2} & -1 \end{array} \right)
\end{equation} 
as long as $0<f<2E$.
Here $\mu=e^{-E/T_{c}}$, $\nu=e^{-E/T_{h}}$,  $a = e^{-f / T_c}$ and $b = e^{-f / T_h}$.
 It is important to note that via the expressions   $p_{1}(t)$, $p_{2}(t)$  and $p_{3}(t) $ that are shown in Appendix A and using the rates,
\begin{eqnarray}
P_{21}&=&{1 \over 2}e^{-(E+f)/T_{c}}, ~P_{12}={1 \over 2},~P_{32}={1 \over 2}e^{-(E+f)/T_{h}}\nonumber \\
 && P_{23}={1 \over 2}, P_{13}={1 \over 2},  P_{31}={1 \over 2} e^{-(2E-f)/T_{c}} 
\end{eqnarray}
 the thermodynamic quantities which are under investigation can be evaluated.  

The net   velocity  $V(t)$ at any time $t$ is the difference between the forward $V_{i}^{+}(t)$ and backward $V_{i}^{-}(t)$ velocities at each site $i$ 
\begin{eqnarray}
V(t)&=& \sum_{i=1}^{3}(V_{i}^{+}(t)-V_{i}^{-}(t)) \\ \nonumber
&=&(p_{1}P_{21}-p_{2}P_{12})+(p_{2}P_{32}-p_{3}P_{23})+\\ \nonumber
&&(p_{3}P_{13}-p_{1}P_{31}).
\end{eqnarray}
At stall force
\begin{eqnarray}
f={E({T_h\over T_c}-1)\over ({2T_h\over T_c}+1)}
\end{eqnarray}
the velocity approaches zero.

Let us next   derive  the   fundamental  entropy relation  
\begin{eqnarray}
S[{p_{i}(t)}]=-\sum_{i=1}^3 p_{i} \ln p_{i}, 
\end{eqnarray}
for the system which is far from equilibrium. Since the hot heat bath located at $i=2$ loses $(E+f)$ amount of heat to the lattice $i=3$ and at the same time gains $(E+f)$ amount of heat from  the lattice $i=3$, one can   write 
the  heat per unit time taken from the hot reservoir is given as 
\begin{eqnarray}
{\dot Q}_{h}(t)&=&(E+f)(p_{2}P_{32}-p_{3}P_{23})\nonumber \\
&=&T_{h} (p_{2}P_{32}-p_{3}P_{23})\ln({P_{32}\over P_{23}})
\end{eqnarray}
as shown in the work \cite{muu17}.
Note that  $\ln({P_{32}\over P_{23}})=(E+f)/T_h$.
On the other hand, the  heat per unit time given to cold reservoir is given by
\begin{eqnarray}
{\dot Q}_{c}(t)&=&(E+f)(p_{2}P_{12}-p_{1}P_{21})+ \nonumber \\
&&(2E-f)(p_{3}P_{13}-p_{1}P_{31}) \nonumber \\
&=&T_{c}(p_{2}P_{12}-p_{1}P_{21})\ln({P_{12}\over P_{21}})+ \nonumber \\
&&T_{c}(p_{3}P_{13}-p_{1}P_{31})\ln({P_{13}\over P_{31}}).
\end{eqnarray}
 Let us  write the entropy extraction (heat dissipation) rate
\begin{eqnarray}
{\dot h}_{d}(t)&=&{-{\dot Q}_{h}(t) \over T_{h}}+{{\dot Q}_{c}(t) \over T_{c}}.
\end{eqnarray}
 Substituting  Eqs. (10) and (11) into Eq. (12) leads to
\begin{eqnarray}
{\dot h}_{d}(t)&=&{-{\dot Q}_{h}(t) \over T_{h}}+{{\dot Q}_{c}(t) \over T_{c}} \nonumber \\
&=&\sum_{i>j}(p_{i}P_{ji}-p_{j}P_{ij}) \ln \left({P_{ji}\over P_{ij}}\right)\nonumber \\
&=&\sum_{i>j}(p_{i}P_{ji}-p_{j}P_{ij}) \ln \left({p_{i}P_{ji}\over p_{j}P_{ij}}\right)-\nonumber \\
&&\sum_{i>j}(p_{i}P_{ji}-p_{j}P_{ij}) \ln \left({p_{i}\over p_{j}}\right) \nonumber \\
&=& {\dot e}_{p}(t)-{\dot S}(t)
\end{eqnarray}
where 
 \begin{eqnarray}
{\dot e}_{p}(t)&=& \sum_{i>j}(p_{i}P_{ji}-p_{j}P_{ij}) \ln \left({p_{i}P_{ji}\over p_{j}P_{ij}}\right)
\end{eqnarray}
and 
\begin{eqnarray}
{\dot S}(t)&=&\sum_{i>j}(p_{i}P_{ji}-p_{j}P_{ij}) \ln \left({p_{i}\over p_{j}}\right).
\end{eqnarray}
Here ${\dot e}_{p}(t)$ and ${\dot S}(t)$ denote the internal  entropy production rate and  the change in the entropy.
{\it From  Eq. (15),   one derives $S[{p_{i}(t)}]=-\sum_{i=1}^3 p_{i} \ln p_{i}$  which implies the fundamental entropy equation is still valid
for the systems that are  driven out of  equilibrium}.

{\it Case2:   Linearly decreasing temperature.\textemdash}
Let us now consider a single  Brownian particle that hops  along one-dimensional discrete ratchet potential (see Eq. (2))   with a load  that  coupled  with a linearly decreasing  temperature   \cite{mu13}
\begin{equation} 
  T_i=T_{h}+{(i-1)(T_{c}-T_{h})\over 3}
   \end{equation}
as shown in Fig. 2.  The indexes $i=1 \cdots 3$.
Once again, the rate equation for the model is given by 
$
{d \vec{p}  \over dt}= {\bold P}\vec{p}
$ 
where $\vec{p}=(p_{1},p_{2},p_{3})^T$.
${\bold P}$ is a 3 by 3 matrix which
is given by
\begin{equation} 
{\bold P}= \left( \begin{array}{ccc}
-\frac{a \mu_1}{2}-\frac{\mu_2^2}{2 a_2}&\frac{1}{2}&\frac{1}{2} \\
\frac{a \mu_1}{2}&\frac{1}{2} (-1-\nu)&\frac{1}{2}\\
\frac{\mu_2^2}{2
a_2}&\frac{\nu}{2}&-1 \end{array} \right)
\end{equation} 
as long as $0<f<2E$.
Here $\mu_1=e^{-E/T_1}$, $\nu=e^{-(E+f)/T_2}$,  $a_1 = e^{-f / T_1}$,  $\mu_2=e^{-E/T_4}$ and $a_2 = e^{-f / T_4}$.
\begin{figure}[ht]
\centering
{
    \includegraphics[width=6cm]{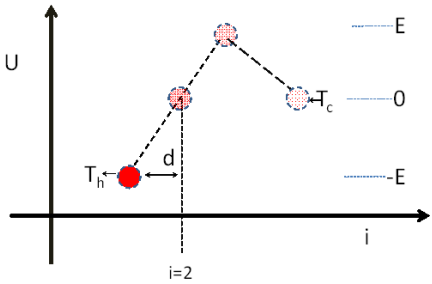}}
\caption{(Color online)  The schematic diagram for a Brownian particle that walks  in a discrete ratchet potential coupled with a linearly decreasing temperature profile. The temperature for the heat baths decreases from $T_h$ to $T_{c}$ according to Eq. (16). } 
\end{figure}
Since  the temperature  linearly  decreases, the parameter $T_1=T_h$, $T_{2}=T_h+(T_c-T_h)/3$, $T_{3}=T_h+2(T_c-T_h)/3$ and $T_4=T_c$. 
The sum of
each column of the matrix ${\bold P}$  is zero, $\sum_{m}{\bold P}_{mn}=0$ which reveals that  the total probability is conserved: $(d/dt)\sum_{n}p_{n}=d/dt({\bold 1}^T\cdot p)={\bold 1}^T\cdot( {\bold P}\vec{p})=0$. It is important to note that via the expressions   $p_{1}(t)$, $p_{2}(t)$  and $p_{3}(t) $ that are shown in Appendix A2 and using the rates,
\begin{eqnarray}
P_{21}&=&{1 \over 2}e^{-(E+f)/T_{1}} ~P_{12}={1 \over 2},~P_{32}={1 \over 2}e^{-(E+f)/T_{2}}\nonumber \\
 && P_{23}={1 \over 2}, P_{13}={1 \over 2},  P_{31}={1 \over 2} e^{-(2E-f)/T_{4}} 
\end{eqnarray}
 the thermodynamic quantities which are under investigation can be evaluated.

Once again, the  velocity  $V(t)$ at any time $t$ is the difference between the forward $V_{i}^{+}(t)$ and backward $V_{i}^{-}(t)$ velocities at each site $i$ 
\begin{eqnarray}
V(t)&=& \sum_{i=1}^{3}(V_{i}^{+}(t)-V_{i}^{-}(t)) \\ \nonumber
&=&(p_{1}P_{21}-p_{2}P_{12})+(p_{2}P_{32}-p_{3}P_{23})+\\ \nonumber
&&(p_{3}P_{13}-p_{1}P_{31}).
\end{eqnarray}
At stall force
\begin{eqnarray}
f={E({T_h-T_c})(4T_h+T_c)\over (2T_h^2+ T_c^2+6T_cT_h)}
\end{eqnarray}
the velocity approaches zero.

The previously derived   relation for the entropy  production rate
\begin{eqnarray}
{\dot e}_{p}(t)&=& \sum_{i>j}(p_{i}P_{ji}-p_{j}P_{ij}) \ln \left({p_{i}P_{ji}\over p_{j}P_{ij}}\right),
\end{eqnarray}
the entropy  extraction rate 
\begin{eqnarray}
{\dot h}_{d}(t)
&=&\sum_{i>j}(p_{i}P_{ji}-p_{j}P_{ij}) \ln \left({P_{ji}\over P_{ij}}\right)\nonumber \\
\end{eqnarray}
and the rate of total entropy 
\begin{eqnarray}
{\dot S}(t)&=&\sum_{i>j}(p_{i}P_{ji}-p_{j}P_{ij}) \ln \left({p_{i}\over p_{j}}\right)
\end{eqnarray}
are still   valid   regardless of any parameter choice. 
For both cases,  as steady state ${\dot e}_{p}(t)={\dot h}_{d}(t)$.
In the absence of load $f$  and in the limit $T_h \to T_c$ (when the system relaxes to its equilibirum state), for both cases, one finds
\begin{eqnarray}
{\dot h}_{d}(t)=0 
\end{eqnarray}
and 
\begin{eqnarray}
 {\dot e}_{p}(t)={\dot S}(t) =-e^{{-3t\over 2}}\ln[{-1+e^{{3t\over 2}}\over 2+e^{{3t\over 2}}}]
\end{eqnarray}
as long as $E=0$. At  stationary state (at equilibirium) $t \to \infty$, ${\dot e}_{p}(t)={\dot h}_{d}(t)={\dot S}(t)=0$.

Because closed-form expressions for 
${\dot S}(t)$, ${\dot e}_{p}(t)$  and ${\dot h}_{d}(t)$ as a function of  $t$ are obtained, 
the analytic expressions for  the change in entropy production,  heat dissipation  and  total entropy can be found analytically via  
\begin{eqnarray}
\Delta h_d(t)&=& \int_{t_0}^{t}\left({\dot h}_{d}(t)\right)dt,
\end{eqnarray}
 \begin{eqnarray}
\Delta e_{p}(t)&=& \int_{t_0}^{t} \left( {\dot e}_{p}(t)  \right)dt 
\end{eqnarray}
and 
\begin{eqnarray}
\Delta S(t) &=&\int_{t_0}^{t} \left({\dot S}(t)\right)dt 
\end{eqnarray}
where $\Delta S(t)=\Delta e_p(t)-\Delta h_d(t)$ and the indexes $i=1 \cdots 3$ and $j=1 \cdots 3$.  
Once again,   in the absence of load $f$,  in the limit $T_h \to T_c$ and when $E=0$, for both cases, one finds
\begin{eqnarray}
\Delta h_d(t)&=& 0,
\end{eqnarray}
\begin{widetext}
\begin{eqnarray}
\Delta S(t)=\Delta e_{p}(t) 
=-\frac{1}{6} \left(-9 t-6 \ln[3]+4 \ln\left[-1+e^{3 t/2}\right]\right)-\frac{1}{6} \left(
+2 \ln\left[2+e^{3 t/2}\right]-4 e^{-3 t/2} \ln\left[1-\frac{3}{2+e^{3
t/2}}\right]\right).
\end{eqnarray}
\end{widetext}
As expected,  in the limit $t \to \infty$, the system approaches  equilibirum state  and  Eq. (30) converges to 
  \begin{eqnarray}
\Delta S=\Delta e_{p}(t)=\ln [3]
\end{eqnarray}
which reconfirm  the  the well known relation for system under infinitesimal process.  In other words,  the system has three  accessible states (three lattices) $\Omega=3$ and  at equilibirium $S=\ln(\Omega)$ .

Furthermore, for  linearly  decreasing  temperature case, the heat dissipation rate is rewritten as 
\begin{eqnarray}
{\dot H}_{d}(t)&=& \sum_{i>j}T_{j}(p_{i}P_{ji}-p_{j}P_{ij}) \ln \left({P_{ji}\over P_{ij}}\right)  \nonumber \\
&=&\sum_{i>j}T_{j}(p_{i}P_{ji}-p_{j}P_{ij}) \ln \left({p_{i}P_{ji}\over p_{j}P_{ij}}\right)-\nonumber \\
&&\sum_{i>j}T_{j}(p_{i}P_{ji}-p_{j}P_{ij}) \ln \left({p_{i}\over p_{j}}\right) \nonumber \\
&=& {\dot E}_{p}(t)-{\dot S}^T(t)
\end{eqnarray}
where 
 \begin{eqnarray}
{\dot E}_{p}(t)&=& \sum_{i>j}T_{j}(p_{i}P_{ji}-p_{j}P_{ij}) \ln \left({p_{i}P_{ji}\over p_{j}P_{ij}}\right)
\end{eqnarray}
and 
\begin{eqnarray}
{\dot S}^T(t)&=&\sum_{i>j}T_{j}(p_{i}P_{ji}-p_{j}P_{ij}) \ln \left({p_{i}\over p_{j}}\right).
\end{eqnarray}
Here the indexes $i=1 \cdots 3$ and $j=1 \cdots 3$.  Our next objective is to write the expression for  the free energy  in terms of  ${\dot E}_{p}(t)$   and  ${\dot H}_{d}(t)$ where  ${\dot E}_{p}(t)$   and  ${\dot H}_{d}(t)$ are the terms that are associated with ${\dot e}_{p}(t)$   and  ${\dot h}_{d}(t)$ except the temperature $T_{j}$.  
Now we have entropy balance equation  ${\dot S}^T(t)={\dot E}_{p}(t)-{\dot H}_{d}(t)$  for our model system.  For  isothermal case and in the absence of load, the system relaxes to its equilibirium. For the case $E=0$,     one finds ${\dot H}_{p}(t)=0$ and 
\begin{eqnarray}
 {\dot E}_{p}(t)={\dot S}^T(t)=T_ce^{{-3t\over 2}}\ln[{-1+e^{{3t\over 2}}\over 2+e^{{3t\over 2}}}].
\end{eqnarray}
In the limit $t \to 0$, ${\dot E}_{p}(t)={\dot S}^T(t)=0$

The second law of thermodynamics  can be   written as  $\Delta S^T(t)=\Delta  E_{p}(t)-\Delta H_{d}(t)$ where $\Delta S^T(t)$, $\Delta  E_{p}(t)$ and $\Delta H_{d}(t)$ are very lengthy expressions which can be evaluated  via 
\begin{eqnarray}
\Delta H_d&=& \int_{t_0}^{t}\left({\dot H}_{d}(t)\right)dt,
\end{eqnarray}
 \begin{eqnarray}
\Delta E_{p}(t)&=& \int_{t_0}^{t} \left({\dot E}_{p}(t) \right)dt
\end{eqnarray}
and 
\begin{eqnarray}
\Delta S^T (t)&=&\int_{t_0}^{t} ({\dot S}^T(t)  )dt.
\end{eqnarray}

 When $f=0$, $T_h \to T_c$ and $E=0$ (system approaching equilibirium), $\Delta H_d=0$  for any time $t$ while 
\begin{widetext}
\begin{eqnarray}
\Delta S^T(t) &=&\Delta E_{p}(t)=-T_c\frac{1}{6} \left(-9 t-6 \ln[3]+4 \ln\left[-1+e^{3 t/2}\right]\right)-T_c\frac{1}{6} \left(
+2 \ln\left[2+e^{3 t/2}\right]-4 e^{-3 t/2} \ln\left[1-\frac{3}{2+e^{3
t/2}}\right]\right).
\end{eqnarray}
\end{widetext}
In long time limit (at equilibirium), $\Delta S(t)=\Delta e_{p}(t)=T_c\ln [3]$.

On the other hand, the total internal energy $U(t)$ is the sum of the internal energies 
\begin{eqnarray}
U[{p_{i}(t)}]&=&\sum_{i=1}^3 p_{i}u_{i} \nonumber \\
&=&p_{1}(t)(-E)+p_{3}(t)(E)
\end{eqnarray}
while the change in the internal energy is given by 
\begin{eqnarray}
\Delta U(t)&=&U[{p_{i}(t)}]-U[{p_{i}(0)}]\nonumber \\
&=&E\left(p_{3}(t)-p_{3}(0)+p_{1}(0)-p_{1}(t)\right).
\end{eqnarray}
We also   verify  the first law  of thermodynamics 
\begin{eqnarray}
{\dot U}[P_{i}(t)]&=&-\sum_{i>j}(p_{i}P_{ji}-p_{j}P_{ij}) \left(u_{i}-u_{j}\right) \nonumber \\
&=&-({\dot H}_{d}(t)+fV(t)).
\end{eqnarray}

Next let us find   the expression for the free energy dissipation rate  ${\dot F}$.  For the isothermal case, the free energy is given by $F=U-TS$ and  next we adapt this relationship to nonisothermal case to write 
\begin{eqnarray}
{\dot F} (t)&=&{\dot U}-{\dot S}^T(t).
\end{eqnarray}  
Substituting Eqs. (32) and (42) in Eq. (43) leads to
\begin{eqnarray}
{\dot F} (t)+{\dot E}_{p}(t)={\dot U }(t)+{\dot H}_{d}(t)=-fV(t)
\end{eqnarray}
which  is the second law of thermodynamics. Note that  
in the absence of load,   ${\dot U }(t)=-{\dot H}_{d}(t)$ and  consequently 
$
{\dot E}_{p}(t)=-{\dot F} (t)
$.   
The change in the free energy  can be written as 
\begin{eqnarray}
\Delta F(t)&=&-\int_{t_0}^{t} \left(  fV(t)+ {\dot E}_{p}(t)   \right)dt \\ \nonumber
        &=&  \int_{t_0}^{t} \left(  {\dot U }(t)+{\dot H}_{d}(t)- {\dot E}_{p}(t)   \right)dt\\ \nonumber
				&=& \Delta U+\Delta H_{d}-\Delta E_{p}.
\end{eqnarray} 
As expected,   at  quasistatic limit where the velocity  approaches  zero  $ V(t)=0$, ${\dot E}_{p}(t) =0$ and ${\dot H}_{d}(t) =0$  and far from quasistatic limit 
$E_{p}>0$  which is  expected as   the engine operates irreversibly. Far from  stall force,   ${\dot E}_{p}(t) \ne {\dot H}_{d}(t)$ as long as a distinct temperature difference between the hot reservoirs is retained. 

For isothermal case, in the absence of potential barrier and load,  $ \Delta U=\Delta H_{d}=0$ and $\Delta F(t)=-\Delta E_{p}$.  At  equilibrium ($t \to \infty$), we get  ${\Delta F}=-T_{c}\ln(3)$. We can recheck this relation via the statistical mechanics approach. In the limit, $t \to \infty$, the probability distributions that are shown in Appendixes A and B, converge 
 \begin{eqnarray}
P_1&=&{1\over 1+ e^{-{E\over T_{c}}}+e^{-{2E\over T_{c}}}} \nonumber \\
P_2&=&{e^{-{E\over T_{c}}}\over 1+ e^{-{E\over T_{c}}}+e^{-{2E\over T_{c}}}}\nonumber \\
P_3&=&{e^{-{2E\over T_{c}}}\over 1+ e^{-{E\over T_{c}}}+e^{-{2E\over T_{c}}}}.
\end{eqnarray}
Accordingly,  the partition function is given by 
\begin{eqnarray}
Z&=&\sum_{i=1}^{3}e^{-{E_{i}\over T_{c}}}\nonumber \\
&=&1+ e^{-{E\over T_{c}}}+e^{-{2E\over T_{c}}}.
\end{eqnarray}
The free energy as well the entropy  can be calculated as $F=-T_{c} \ln(Z)$ and $S=\ln(z)+\bar{E} \beta$. When $E \to 0$, the free energy converges to $F=-T_{c} \ln(3)$  while the entropy approaches $S= \ln(3)$.

\section{ Entropy production rate and  Free energy}

Hereafter, whenever we plot  the  figures, we use  dimensionless quantities $\epsilon=E/T_{c}$, $\lambda=f/T_{c}$ and $\tau={T_{h}\over T_{c}}$. We also introduce dimensionless time ${\bar t}=\Gamma t$ and after this the bar will be dropped.

{\it Entropy.\textemdash} The dependence of entropy  on the  model parameters can be explored  via Eq. (9).   As shown  in Fig. 3,  the  entropy of the system  exhibits an intriguing parameter dependence.  
\begin{figure}[ht]
\centering
{
    \includegraphics[width=6cm]{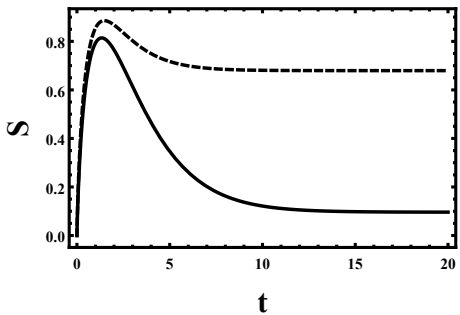}
}
\caption{ (Color online)  The entropy ${ S}(t)$  as a function of  $t$   evaluated analytically via Eq. (9) for a given  $\epsilon=4.0$, $f=0.0$ and   $\tau=2.0$. The dashed line indicates the plot for   a heat bath where its temperature linearly decreases along with the reaction coordinate while the solid line is  plotted by considering a  Brownian particle   that operates between the hot and cold baths.     } 
\label{fig:sub} 
\end{figure}
As shown in the figure, for $t \ne 0$, $S>0$ which indicates that in the presence of symmetry-breaking fields such as nonuniform temperature or external force, the system is driven out of equilibrium.  ${ S}(t)$ is also considerably larger for linearly decreasing temperature case than the entropy for  Brownian particle that operates between the hot and cold baths. This suggests that that the entropy production is higher for the system that operates in a heat bath where its temperature decreases linearly along with the reaction coordinate.  For isothermal case $T_{h} = T_{c}$ as well as in the absence of both external load and bistable potential $U_0=0$, the particle undergoes a random walk on a lattice. For both cases, $S(t)$  converges to  $S(t)\to \ln[3]$ in the limit  $t\to \infty$.

\begin{figure}[ht]
\centering
{
    \includegraphics[width=6cm]{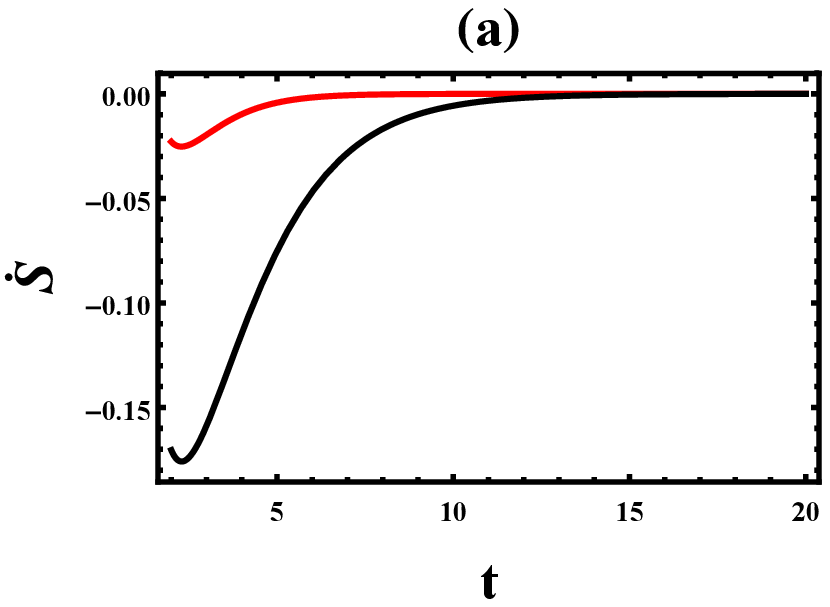}}
\hspace{1cm}
{
    \includegraphics[width=6cm]{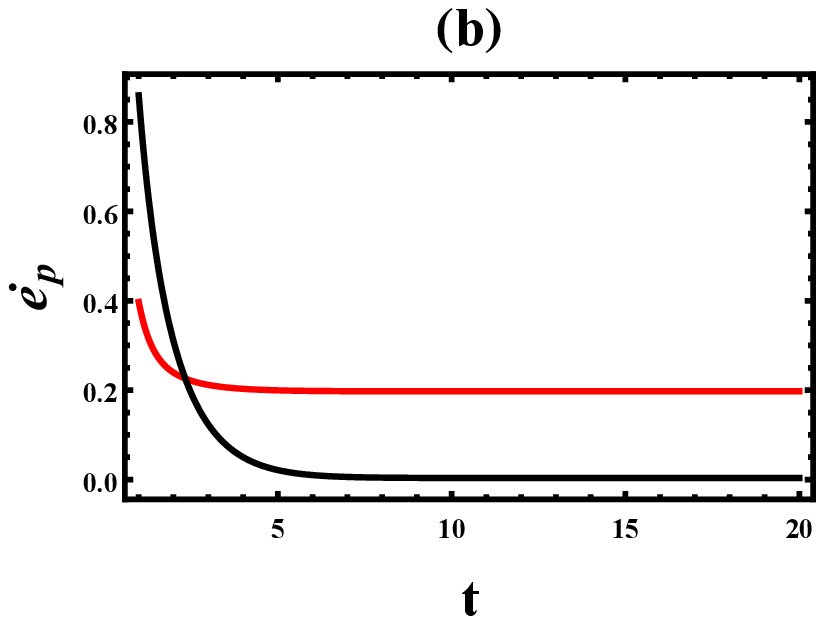}
}
{
    \includegraphics[width=6cm]{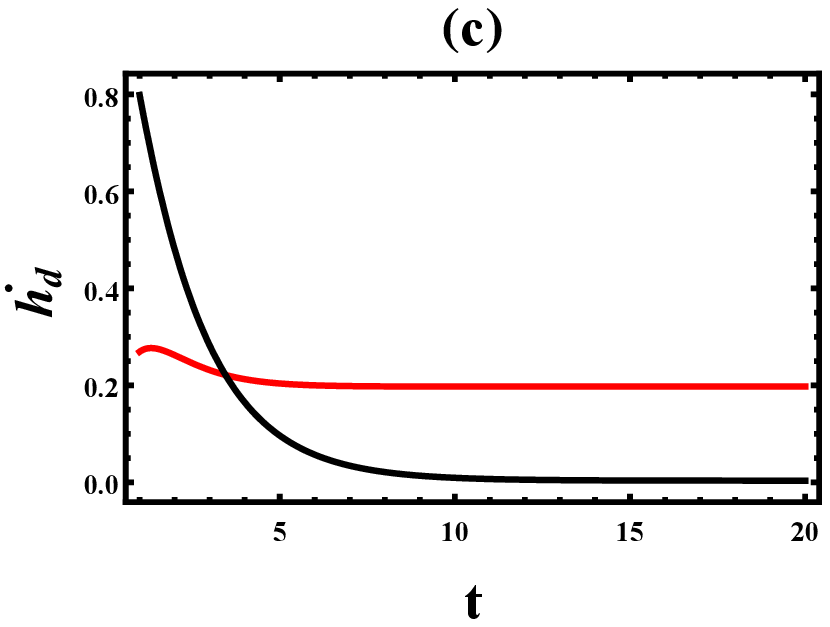}
}
\caption{ (Color online) (a)  ${\dot S}(t)$ versus $t$ is evaluated analytically   via Eqs. (15) and (23).   (b) The entropy production rate ${\dot e}_{p}(t)$  as a function
of $t$. ${\dot e}_{p}(t)$  is analyzed analytically via Eqs. (14) and (21). (c) The entropy extraction rate  ${\dot h}_{d}(t)$ as a function of
$t$ evaluated analytically using  Eqs. (13) and (22). 
In the figures, the red line indicates the plot for a heat bath where its temperature linearly decreases while the black solid line is plotted by considering  a  Brownian particle that operates between the hot and cold baths. Clearly  ${\dot h}_{d}(t)$ and  ${\dot e}_{p}(t)$ are considerably large for linearly decreasing temperature case. In the figures, we fix    $\epsilon=2,0$,  $\tau=20.0$,  $\lambda=0.6$. 
 } 
\label{fig:sub} 
\end{figure}

\begin{figure}[ht]
\centering
{
    \includegraphics[width=6cm]{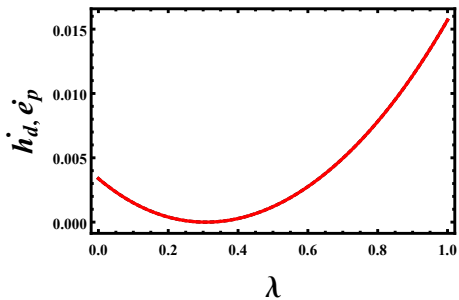}
}
\caption{ (Color online) The entropy production rate ${\dot e}_{p}(t)$   and the entropy extraction rate  ${\dot h}_{d}(t)$   versus  $\lambda$. The entropy production rate is analized  via  Eqs. (14) and (21) while the entropy extraction rate
is evaluated analytically using  Eqs. (13) and (22). In the figure, the parameters are fixed as  $\epsilon=2$,  $\tau=2.0$ and $t=10^6$ (steady state).  
    At steady state  ${\dot e}_{p}(t)={\dot h}_{d}(t)$.  } 
\label{fig:sub} 
\end{figure}

{\it Entropy production rate.\textemdash} Next let us explore the dependence for the rate of entropy production ${\dot e}_{p}(t)$, the rate of entropy  ${\dot S}(t)$ and the rate of entropy flow from the system to the outside  ${\dot h}_{d}(t)$ on the system parameters. 
The expression for ${\dot S}(t)$,  ${\dot h}_{d}(t)$  and  ${\dot e}_{p}(t)$  can be evaluated via Eqs. (23), (22), and (21), respectively as shown in Fig. 4. In the figure, we plot ${\dot S}(t)$,  ${\dot h}_{d}(t)$  and  ${\dot e}_{p}(t)$ as a function of $t$.   In the figure, the red line indicates the plot for a heat bath where its temperature linearly decreases while the solid line is plotted by considering a  Brownian particle that operates between the hot and cold baths. The fact that ${\dot e}_{p}(t)>0$ and ${\dot h}_{d}(t)>0$  exhibits that the system is exposed to symmetry-breaking fields
such as external force or nonuniform temperature.  As a result, the system is driven out of equilibrium.  The entropy,   entropy production, and extraction rates are also  higher for linearly decreasing cases than the particle that operates between two heat baths indicating that a system that operates in a heat bath where its temperature decreases linearly with the reaction coordinate exhibits a  higher level of irreversibility.  
As expected,  ${\dot e}_{p}(t)$ and ${\dot h}_{d}(t)$ approach their steady-state values  ${\dot e}_{p}= {\dot h}_{d}$ as time progresses. Even in the absence of symmetry-breaking fields, as long as the system is operating in a finite time, the system exhibits irreversible dynamics and as a result ${\dot e}_{p}>0$ for small $t$ and decreases  (the system relaxes to equilibrium) as time increases. 
In the limit $t \to \infty$, ${\dot e}_{p}={\dot h}_{d}=0$.

The velocity approaches zero (quasistatic limit) in the vicinity of a stall force or when $U_0 \to 0$.   
At quasistatic limit,  regardless of any parameter choice, we find ${\dot e}_{p}={\dot h}_{d}(t)=0$. One should note that  the vanishing of velocity may not indicate the  system is at 
thermodynamic equilibrium as  pointed out  by   Ge $et.$ $al.$ \cite{mg22}.
This can be appreciated by plotting ${\dot e}_{p}(t)$ or ${\dot h}_{d}(t)$ (using Eqs. 21 and 22) as a function of load  in long time limit.   For  linearly decreasing temperature case,  we plot ${\dot e}_{p}(t)$ and ${\dot h}_{d}(t)$ as  a function of load  in Fig. 5.  The figure depicts that  in the long time limit,  both ${\dot e}_{p}(t)$ and ${\dot h}_{d}(t)$  attain a zero value at stall force.  
\begin{figure}[ht]
\centering
{
    \includegraphics[width=6cm]{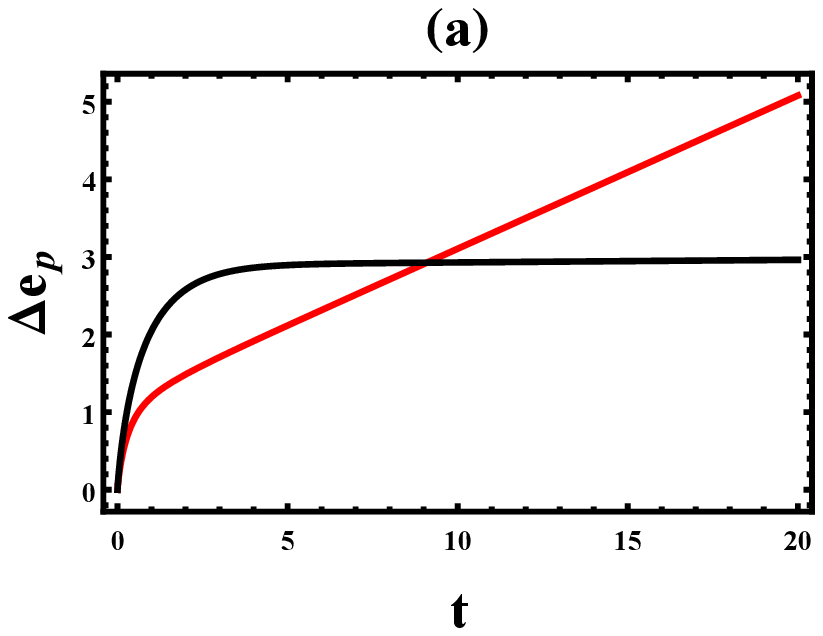}}
\hspace{1cm}
{
    \includegraphics[width=6cm]{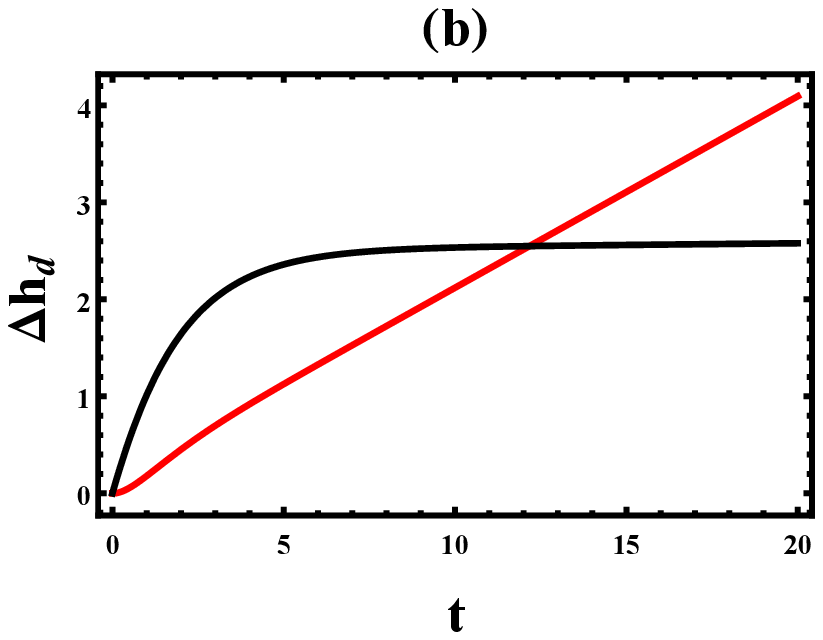}
}
\caption{ (Color online) (a)  $\Delta e_p(t)$   as a function of  $t$  that  evaluated analytically via Eq. (27)  for fixed $\epsilon=2.0$,  $\tau=20.0$ and $\lambda=0.6$.   (b)   ${\Delta  h}_{d}(t)$  as a function of  $t$  is plotted  using Eq. (26) for fixed $\epsilon=2.0$,  $\tau=20.0$ and $\lambda=0.6$. In the figures, the  red line  shows the plot  for a linearly decreasing temperature case and the black line is plotted  by considering  a  Brownian particle that operates between the hot and cold baths.   } 
\label{fig:sub} 
\end{figure}

The dependence  for  
$\Delta S(t)$ , $\Delta e_p(t)$ and $\Delta h_d(t)$ as a function of  $t$  can be  explored  employing Eq. (28), (27) and (26) respectively. 
The expressions for $\Delta h_d(t)$,  $\Delta S(t)$  and $\Delta e_p(t)$  are lengthy and will not be presented in this work. As shown in Figs. 6a and 6b, for a system that operates between hot and cold reservoirs,  $\Delta h_d(t)$  and $\Delta e_p(t)$ approach a non-equilibrium steady state in the long time limit.
 However, for a system that operates in heat baths where its temperature decreases linearly, $\Delta S(t)$  and $\Delta e_p(t)$ increase linearly as time progresses. This reveals that, unlike systems that operate between hot and cold reservoirs,  this system exhibits a higher level of irreversibility. Next we explore   the dependence of  the free energy dissipation rate shown in Eqs. (43) and (44) on $t$. In general ${\dot F}<0$ and approaches  zero in the long time limit for both cases (see Fig. 7).

\begin{figure}[ht]
\centering
{
    \includegraphics[width=6cm]{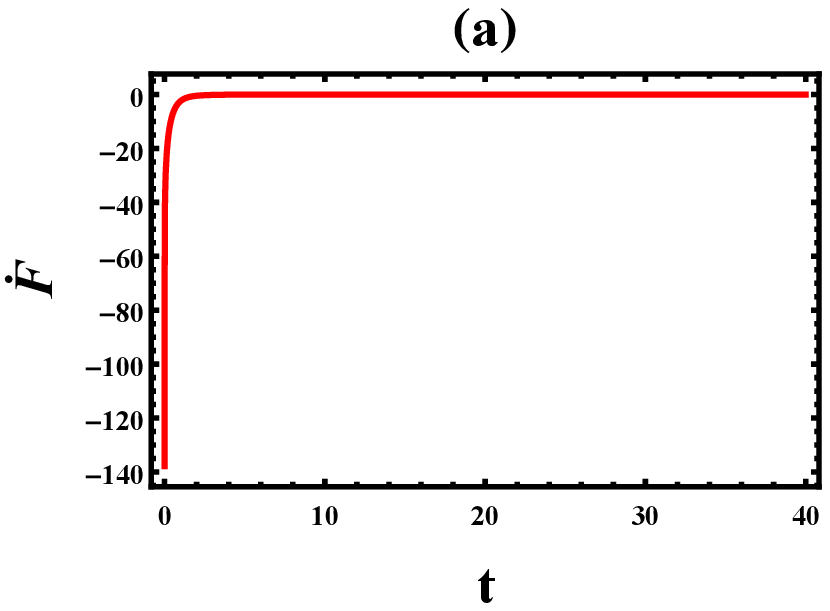}}
\hspace{1cm}
{
    \includegraphics[width=6cm]{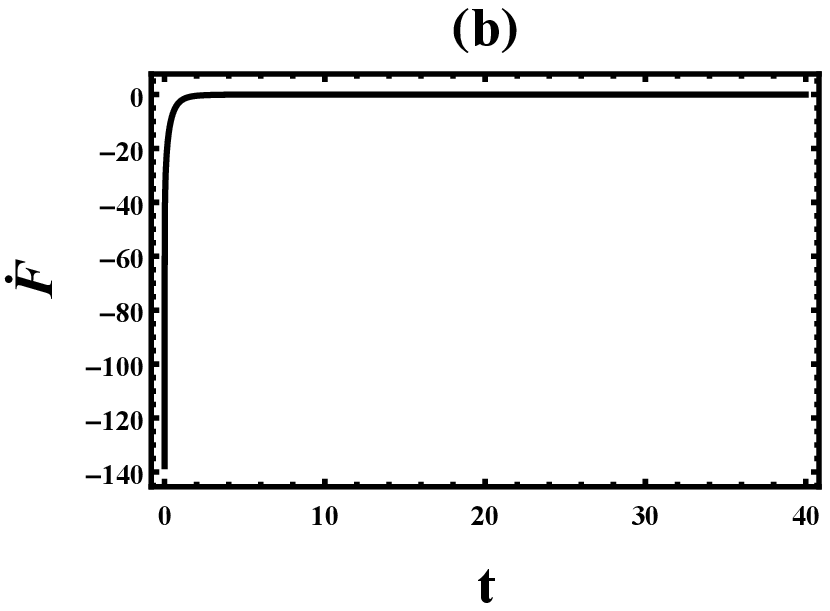}}
\caption{ (Color online) (a) The free energy dissipation rate  ${\dot F}$  versus $t$  for a linearly decreasing temperature  case is plotted using Eq. (43).   (b) The free energy dissipation rate  ${\dot F}$  as a function of  $t$ for a heat bath that coupled with the hot and cold temperature employing Eqs. (43) and (44). For both figures, the parameters  are fixed  as $\epsilon=2.0$,  $\tau=20.0$ and $\lambda=0.6$.   }
\label{fig:sub} 
\end{figure}

\begin{figure}[ht]
\centering
{
    \includegraphics[width=6cm]{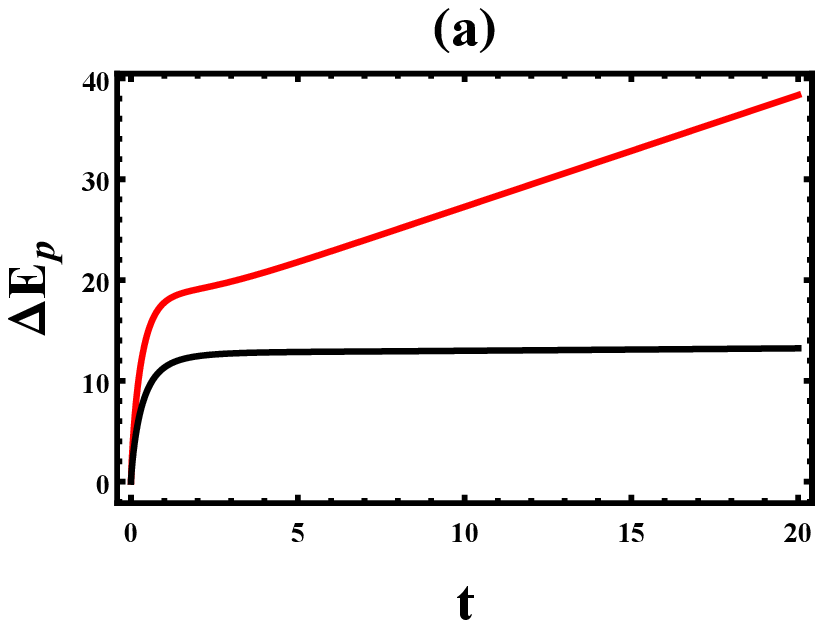}}
\hspace{1cm}
{
    \includegraphics[width=6cm]{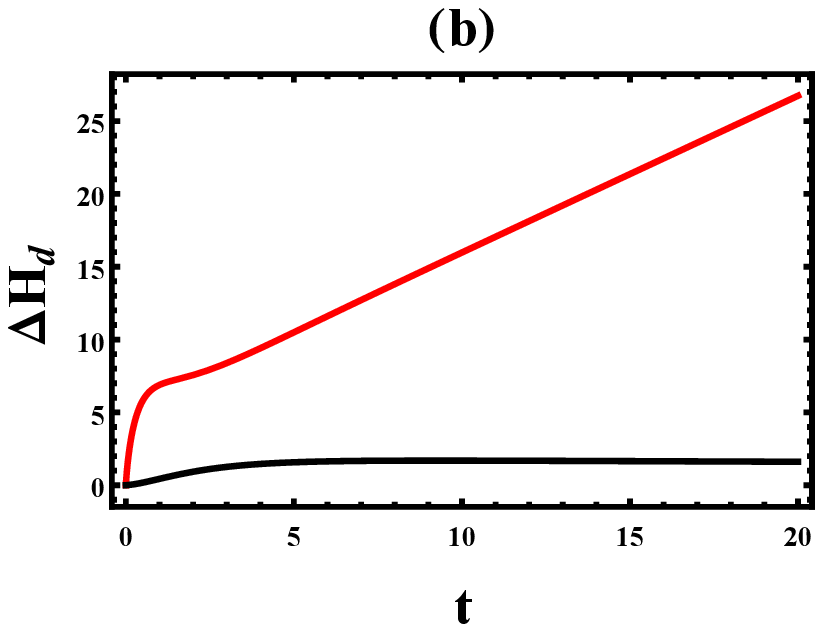}
}
\caption{ (Color online)  (a)  $\Delta E_p(t)$   as a function of  $t$ is evaluated via Eq. (37).    (b) The  plot of    ${\Delta  H}_{d}(t)$  versus    $t$ is plotted employing Eq. (36). In the figures, the parameters are  fixed as $\epsilon=2.0$,  $\tau=20.0$ and $\lambda=0.6$.     For both figures, the  red line  shows the plot  for a linearly decreasing temperature case and the black line is plotted  by considering  a  Brownian particle that operates between the hot and cold baths.    }
\label{fig:sub} 
\end{figure}

As discussed before,  once  the  expressions for   ${\dot H}_{d}(t)$,   ${\dot E}_{p}$, ${\dot S}^T$   are analyzed, the corresponding entropy balance equation can be calculated  as  $
{d S^T(t)\over dt}={\dot E}_{p}-{\dot H}_{d}$. The expressions for these relations are very complicated. In Fig. 8, using Eqs. (36) and (37), we plot   ${\Delta  H}_{d}(t)$  and $\Delta E_p(t)$ as a function of $t$ for fixed $\epsilon=2.0$,  $\tau=20.0$ and $\lambda=0.6$.  In the figure,  the red line shows the plot for a linearly decreasing temperature case, and the black line is plotted by considering a  Brownian particle that operates between the hot and cold baths. 
Surprisingly, although $\Delta E_p(t)$   and ${\Delta  H}_{d}(t)$  saturates to a constant value for a heat engine that operates between the hot and cold heat baths,  for linearly decreasing temperature case both  $\Delta E_p(t)$   and ${\Delta  H}_{d}(t)$  step up in time linearly. This justifies that, unlike systems that operate between hot and cold reservoirs,  systems that operate in heat baths where their temperature decreases linearly have a higher level of irreversibility.

\begin{figure}[ht]
\centering
{
    \includegraphics[width=6cm]{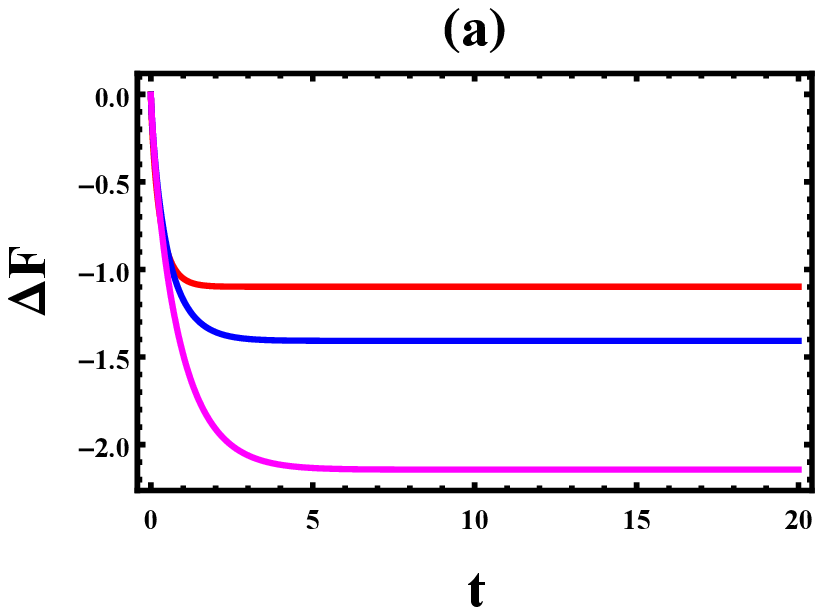}}
\hspace{1cm}
{
    \includegraphics[width=6cm]{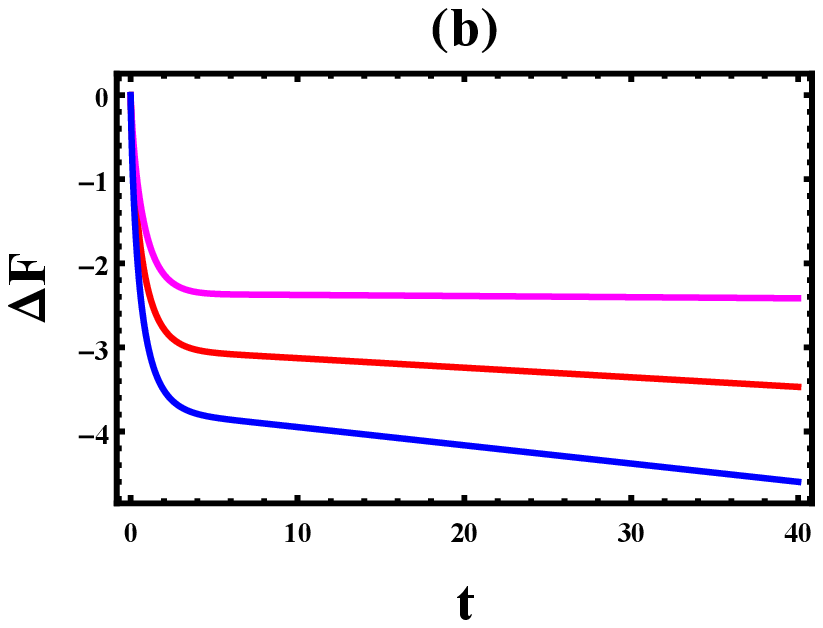}}
\caption{ (Color online) (a)  
The change in free energy   ${\Delta F}$  versus $t$  is plotted using Eq. (45) for the system that operates between the hot and cold baths.  The figure depicts that at a steady-state,  the free energy saturates to a constant but minimal value.  (b) Fig. 9b  shows the dependence 
the change in free energy   ${\Delta F}$  versus $t$ for the system that operates in a linearly decreasing temperature profile.  The figure depicts that as time progresses, the free energy decreases linearly. In both figures, the load  is  fixed as   $\lambda=0.2$. The barrier height is also fixed as $\epsilon=0$, $\epsilon=1.0$ and $\epsilon=2.0$  from the top to bottom. 
  } 
\label{fig:sub} 
\end{figure}

Exploiting Eq. (45), let us now investigate further how the free energy behaves as a function of the system parameters.  Fig. 9a  depicts the plot for 
the change in free energy   ${\Delta F}$  versus $t$ for the system that operates between the hot and cold baths.    On the other hand,  Fig. 9b  shows the plot for 
the change in free energy   ${\Delta F}$  versus $t$ for the system that operates in a linearly decreasing temperature profile. The fact that ${\Delta F} \ne 0$,  indicates that our model system is inherently irreversible even within the long time limit.
The change in free energy   ${\Delta F}$  decreases in time and saturates to a constant but minimal value for the system that operates between the hot and cold baths.  On the contrary, for the system that operates in a linearly decreasing temperature case,  the change in free energy decreases linearly indicating that the degree of irreversibility is higher for such systems.

\section{The efficiency and velocity of the heat engine   }

As discussed before,  in the presence of  external force,   the velocity approaches  zero  $ V(t)=0$ in the vicinity of the stall force. For the Brownian heat engine that operates between  two heat baths,  the stall  force is given as 
\begin{equation}
f={E ({T_{h}\over T_{c}}-1)\over (2 {T_{h}\over T_{c}}+1)}.
\end{equation}
while for linearly   decreasing  thermal arrangement case, the stall force is calculated as  
\begin{eqnarray}
f={E({T_h-T_c})(4T_h+T_c)\over (2T_h^2+ T_c^2+6T_cT_h)}.
\end{eqnarray}
Evaluating  $E_{p}$ near the stall force, one finds  $E_{p}=0$ as long as the system is at a  steady-state regime. Far from a steady-state regime (even in the vicinity of  the stall force), 
$E_{p}>0$  which is expected as the engine operates irreversibly. 
 For isothermal case  without load, $E_{p}=0$ at stationary state.  Next, we explore the dependence of the velocity and  efficiency on the system parameters.

\subsection{ The particle's velocity}
 
The  velocity  of the particle   is  sensitive to time. Our analysis indicates that the velocity of the particle depends on the system parameters. For  instance,  the velocity of the particle is positive  when $T_{h} \ne T_{c}$ and $f=0$. For isothermal case $V<0$ as long as $f>0$. In general for $T_{h}\ne T_{c}$ and $f>0$, the system exhibits fascinating dynamics where $V>0$ when the load is less than the stall force $f'$, $f<f'$ and $V<0$ if $f>f'$. This suggests that the mobility of the particle can be manipulated by varying the external force.

\begin{figure}[ht]
\centering
{
    \includegraphics[width=6cm]{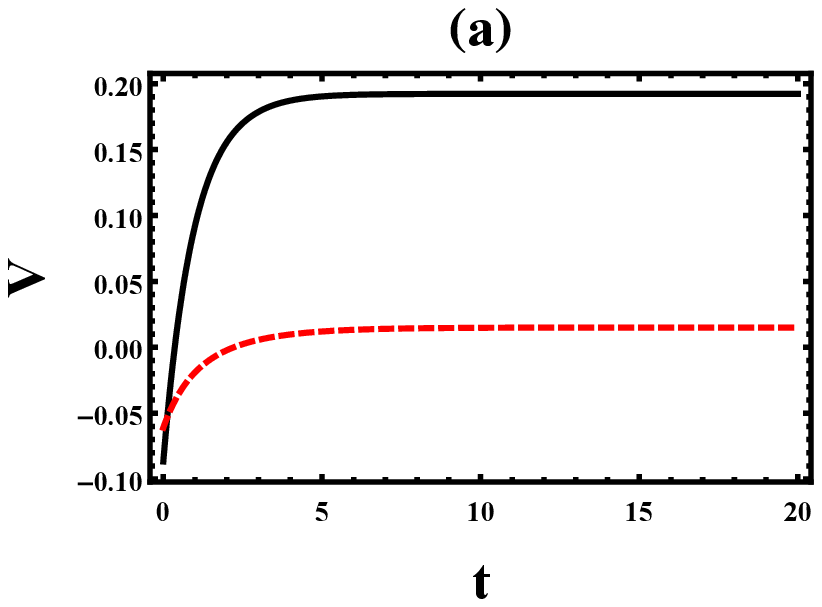}}
\hspace{1cm}
{
    \includegraphics[width=6cm]{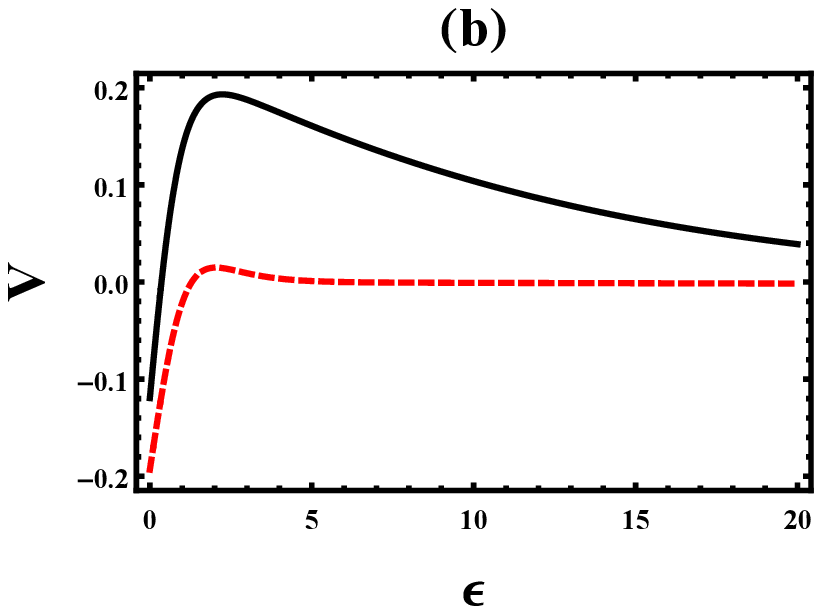}
}
\caption{ (Color online) (a) Particle velocity  $V$  as a function of  $t$  is plotted  via Eqs. (7) or (19) for fixed $\epsilon=2.0$,  $\tau=20.0$ and $\lambda=0.6$  (b)  Particle velocity  $V$  as a function of $\epsilon$ is analized  employing via Eqs. (7) or (19) for fixed $t=10.0$,  $\tau=20.0$ and $\lambda=0.6$. In both  figures, the black solid line indicates the plot for a heat bath where its temperature linearly decreases while the red  line is plotted by considering  a  Brownian particle that operates between the hot and cold baths.   }
\label{fig:sub} 
\end{figure}

\begin{figure}[ht]
\centering
{
    \includegraphics[width=6cm]{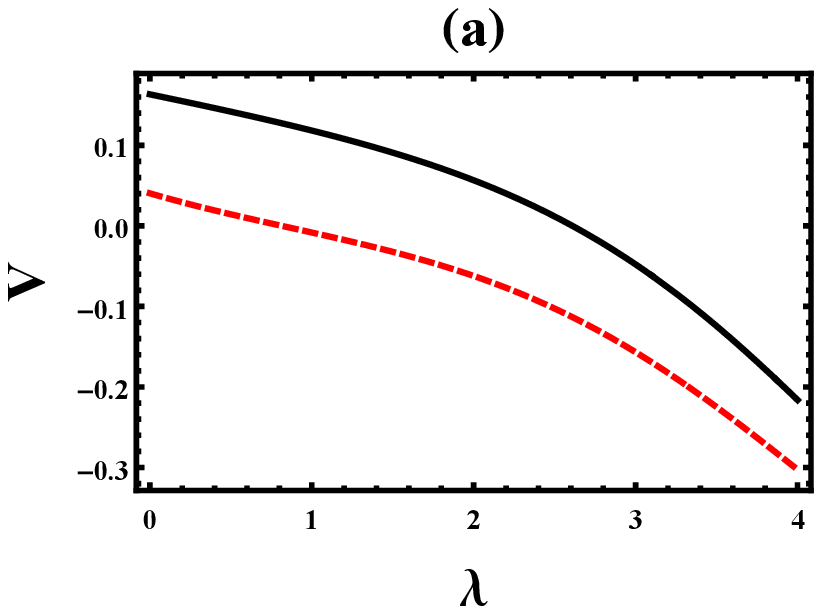}}
\hspace{1cm}
{
    \includegraphics[width=6cm]{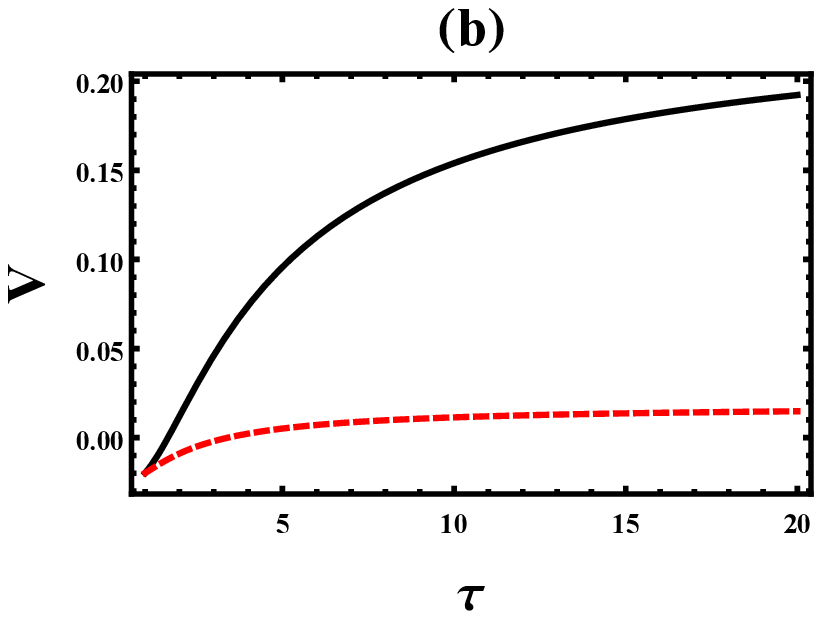}
}
\caption{ (Color online)  (a)  The particle velocity $V$  as a function of  $\lambda$  is   evaluated via Eqs. (7) or (19) for fixed $\epsilon=2.0$,  $\tau=8.0$ and $t=10.0$. (b)  The particle velocity    $V$  as a function of $\tau$ is plotted employing  Eqs. (7) or (19)   for fixed $t=10.0$,  $\epsilon=2.0$ and $\lambda=0.6$. In   figures, the black solid line indicates the plot for a heat bath where its temperature linearly decreases while the red  line is plotted by considering  a  Brownian particle that operates between the hot and cold baths.   }
\label{fig:sub} 
\end{figure}

The dependence of the velocity on time is also explored via Eqs. (7) or (19).  The time $t$ dictates the magnitude and the direction of the velocity.  This can be appreciated by plotting $V$ as a function of time (see  Fig. (10a)). Figure 10a depicts that for small $t$,  the net particle flow is in the reverse direction  (negative).   As time increases, the magnitude of $V$ increases and saturates to a constant value. The particle velocity is significantly higher for the  Brownian particle that operates in the thermal bath where its temperature decreases linearly than a Brownian particle that operates between two heat baths. In the figure, we fix $\epsilon=2.0$,  $\tau=20.0$, and $\lambda=0.6$. In both  figures, the black solid line indicates the plot for a heat bath where its temperature linearly decreases while the red  line is plotted by considering  a  Brownian particle that operates between the hot and cold baths.

Exploting  Eqs. (7) or (19) further,  in Fig. 10b  we  plot  the velocity  $V$ as a function of $\epsilon$.  The figure depicts that the particle manifests a peak velocity at a particular barrier height $\epsilon^{max}$ and at this particular height, the engine
operates with maximum power.  Once again the particle velocity is considerably higher for the  Brownian particle that operates in the thermal bath where its temperature decreases linearly than a Brownian particle that operates between two heat baths. In the figure,  we fix  $t=10.0$,  $\tau=20.0$ and $\lambda=0.6$.

The dependence for the  velocity $V$  on  load  is explored employing Eq. (19) as shown in Fig. 11. In the figure, we fix  $t=10.0$,  $\epsilon=2.0$ and $\lambda=0.6$. The figure depicts that as long as the load is less than the stall force, $V>0$ while when the load is greater than the stall force, $V<0$.  At stall force, the particle velocity becomes zero. On the other hand,  the velocity steps up as the $\tau$ increases as depicted in  Fig. 11b.

\subsection{The efficiency of the heat engine}
Let us now explore how the efficiency $\eta$  behaves as the model parameters vary. For both cases, the rate of work done is given as  ${\dot W}=fV(t)$. On the contrary, the rate of heat input is model dependent. For  the case where the system  operates between the two baths, the rate of heat input is given as ${\dot Q}_{in}(t)={\dot Q}_{h}(t)=T_{h} (p_{2}P_{32}-p_{3}P_{23})\ln({P_{32}\over P_{23}})$. For linearly decreasing temperature case, since the heat bath 
from the left potential well contributes  for the particle to jump to the right,  the particle  must get   ${\dot Q}_{in}(t)=T_{h} (p_{2}P_{32}-p_{3}P_{23})\ln({P_{32}\over P_{23}})+T_{c}(p_{2}P_{12}-p_{1}P_{21})\ln({P_{12}\over P_{21}})$ amount of heat  from the system.  The efficiency then is given by
\begin{eqnarray}
\eta= {{\dot W}\over{\dot Q}_{in}(t)}.
\end{eqnarray}
In general, the efficiency of the system increases in time and at a steady state, the system attains maximum efficiency. The efficiency  at the quasistatic limit  can be obtained  via  Eq. (50).  For a Brownian heat engine that operates between two heat baths, one gets 
\begin{eqnarray}
\eta= 1-{T_{c}\over T_{h}}
\end{eqnarray}
which is  the efficiency of the Carnot heat engine. For the heat engine that operates  in a heat bath that decreases linearly, at the quasistatic limit, we get  
\begin{eqnarray}
\eta= 1 -{ (T_c (T_c + 5 T_h))\over (2 T_h (T_c + 2 T_h))}
\end{eqnarray}
which is approximately equal to the 
 efficiency of the endorevesible heat engine $\eta_{CA}$ 
  \begin{equation}   
\eta_{CA}=1-\sqrt{T-c/T_h}
\end{equation}   
as long as  the temperature difference between the hot and the cold reservoirs is not large. 
 This can be further  appreciated by Taylor expanding  Eqs. (52) and (53)  around $\tau=1$.
 As discussed in the work \cite{mu25}, it is still unknown why different model systems approach the Taylor expression shown above.  

The dependence of the efficiency $\eta$   on the model parameters is also  explored
 by omitting the heat exchange
via kinetic energy. The efficiency  $\eta$ as a function of rescaled load  is evaluated  analytically  (see Fig. 12a) via Eq. (50). The figure is plotted by fixing   $\epsilon=2.0$, $\tau=2.0$ and $t=1000.0$.  The figure exhibits that $\eta$ increases to its  maximum (quasistatic limit) value.  The efficiency is considerably large for the system that operates between two heat baths.   
	The efficiency  $\eta$ as a function of barrier height is plotted in Fig. 12b for the  parameter values of  $\lambda=0.2$, $\tau=2.0$ and $t=1000.0$. The efficiency decreases as the barrier height increases.  When the magnitude of the rescaled temperature steps up, the efficiency of the system monotonously increases.

\begin{figure}[ht]
\centering
{
    \includegraphics[width=6cm]{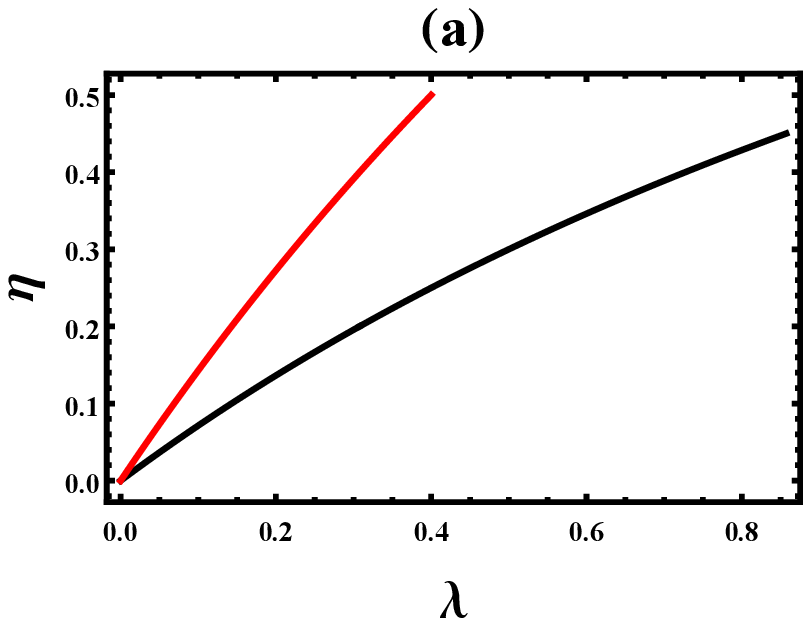}}
\hspace{1cm}
{
    \includegraphics[width=6cm]{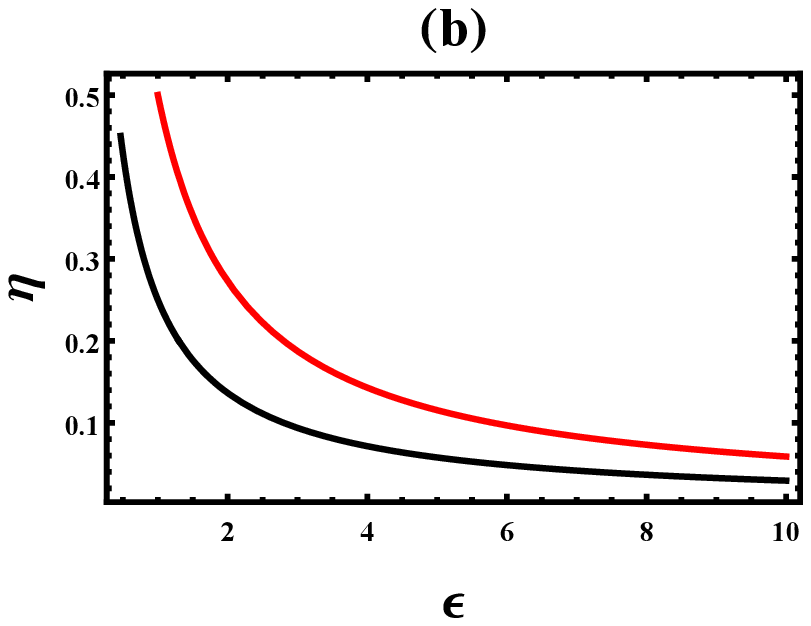}
}
\caption{ (Color online) (a) The efficiency   $\eta$  as a function of  $\lambda$  evaluated using Eq. (50) for fixed values of  $\epsilon=2$,  $\tau=2.0$ and $t=1000.0$.  (b)  The efficiency $\eta$  as a function of $\epsilon$  is plotted  employing Eq. (50) for fixed values of $t=1000.0$,  $\tau=2.0$ and $\lambda=0.2$. The solid line  represents the plot for linearly decreasing temperature case while the  red dashed line  is plotted for a heat engine that operates between the hot and cold baths.    }
\label{fig:sub} 
\end{figure}

\section{Summary and conclusion}

Studying the thermodynamic feature of nonequilibrium systems is challenging since their energy as well as the particles' flux constantly changes in time. Consequently,  exploring the thermodynamics feature of non-equilibrium systems requires a more general concept as well as rigorous
mathematical analysis.  Due to the  lack of exact solutions, most of the previous works  addressed how 
different thermodynamics    features 
behave either at the quasistatic limit or at steady-state regimes. To fill this gap, in this work, we present an exactly solvable model that helps to explore the thermodynamic features of systems beyond a linear response and steady-state regime.  Not only the long-time property (steady-state) but also the short-time behavior of the system is explored by obtaining exact time-dependent solutions.  The general expressions for free energy, entropy production as well as entropy extraction rates are derived for a system that is genuinely driven out of equilibrium by time-independent force as well as by spatially varying thermal background.

From an equilibrium thermodynamics point of view, the entropy $S(t)$  is the most 
explored physical quantity.  It is a well-known fact that even in the absence of  symmetry-breaking  fields, 
the entropy of systems can be greater than zero $S(t)>0$  as long as the system operates in a finite time and only in a long time limit does the system become reversible $S(t)=0$. However, in the presence of symmetry-breaking fields,   the systems are driven out of equilibrium even within the long time limit. In this regard,  most of the previous studies focused on exploring how the entropy $S(t)$,  the production ${\dot e}_{p}(t)$ and extraction rates ${\dot h}_{d}(t)$  behave either at steady state of quasistatic limit.  To comprehend the thermodynamic features
of systems beyond a linear response and steady-state
the regime, we solve the model system analytically.  The thermodynamic properties of a system that operates between the hot and cold baths are also compared and contrasted ( as a function of time) with a system that operates in a heat bath where its temperature linearly decreases along with the reaction coordinate. For our model system, the fact that the entropy, 
the entropy production, and extraction rate are greater than zero suggests that in the presence of symmetry-breaking fields such as nonuniform temperature or external force, the system is driven out of equilibrium. The entropy production ${\dot e}_{p}(t)$ and extraction rates ${\dot h}_{d}(t)$ 
are also considerably larger for linearly decreasing temperature case than the entropy for  Brownian particle that operates between the hot and cold baths. This suggests that the degree of irreversibility is higher for the system that operates in a heat bath where its temperature decreases linearly along with the reaction coordinate.

According to the celebrated statistical thermodynamics theory,   systems tend to maximize their entropy or minimize their free energy in the effort to reach thermal equilibrium, and consequently, the change in free energy  ${\Delta F} = 0$.  This also implies for any irreversible system, the free energy difference is a positive thermodynamic quantity.  To discern the thermodynamic
features of systems beyond equilibrium regime,  we further explore the dependence of free energy on the system parameters for a system that is genuinely driven out of equilibrium. For our system,  ${\Delta F} \ne 0$ and this indicates that our model system is inherently irreversible even within the long time limit. The change in free energy   ${\Delta F}$  decreases in time and saturates to a constant but minimal value for the system that operates between the hot and cold baths.  On the contrary, for the system that operates in a linearly decreasing temperature case,  the change in free energy decreases linearly indicating that the degree of irreversibility is higher for such a system.  
The term that is related to entropy production rate ($\Delta E_p(t)$)   and the heat dissipation rate (${\Delta  H}_{d}(t)$)  saturates to a constant value for a heat engine that operates between the hot and cold heat baths. Surprisingly for the linearly decreasing temperature case both  $\Delta E_p(t)$   and ${\Delta  H}_{d}(t)$  step up in time linearly. This justifies that, unlike systems that operate between hot and cold reservoirs,  systems that operate in a heat bath where its temperature decreases linearly have a higher level of irreversibility.

The energetics of the system that operates between the hot and cold baths are also compared and contrasted with a system that operates in a heat bath where its temperature linearly decreases along the reaction coordinate. We show that a system that operates between the hot and cold baths has significantly lower velocity but a higher efficiency in comparison with a linearly decreasing case.  For a  linearly decreasing background temperature case,  we show that the efficiency of such a Brownian heat engine is lower than Carnot's efficiency even at the quasistatic limit.  At the quasistatic limit, the efficiency of the heat engine approaches the efficiency of the endoreversible engine.

Due to the lack of exact analytic results, most of the previous studies studied  
the thermodynamic features of systems in a linear response and steady-state regime. The exactly solvable model presented in this work enables us to explore how the free energy, total entropy, entropy production, and extraction rates behave as a function of time far beyond linear response and steady-state regime. The change in free energy in particular exhibits an elegant time dependence. Only for a linearly decreasing thermal arrangement does the free energy monotonously decrease 
in time revealing the way how the temperature is arranged in the reaction coordinate affects the free energy.   In conclusion,  even though a specific model system is considered, the thermodynamic relations that are obtained in this work are generic and vital to advance the nonequilibrium statistical mechanics.

\section*{ Appendix A1}
In this Appendix we will give the expressions for $p_{1}(t)$, $p_{2}(t)$  and $p_{3}(t)$ as well as $V(t)$ for a  Brownian particle   that operates between the hot and cold baths. 
For the particle which is initially  situated at  site $i=1$,  the time dependent  normalized probability distributions after solving the rate equation ${d \vec{p}  \over dt}= {\bold P}\vec{p}$ are calculated as 
\begin{eqnarray}
p_{1}(t)&=&c_{1}\frac{a (2+\nu b)}{\mu \left(\mu+\left(a^2+\mu\right) \nu b\right)}+\\ \nonumber
& &c_{2} e^{-\frac{\left(a+a^2 \mu+\mu^2\right) t}{2 a}} \left(-1+\frac{a
(-1+a \mu)}{-\mu^2+a \nu b}\right),\\
p_{2}(t)&=&-c_{3} e^{\frac{1}{2} t (-2-\nu b)}-c_{2}\frac{a\text{  }e^{-\frac{\left(a+a^2 \mu+\mu^2\right) t}{2 a}} (-1+a \mu)}{-\mu^2+a
\nu b}+\\ \nonumber
&&c_{1}\frac{ \left(2 a^2+\mu\right)}{\mu+\left(a^2+\mu\right) \nu b},\\
p_{3}(t)&=&c_{1}+c_{2} e^{-\frac{\left(a+a^2 \mu+\mu^2\right) t}{2 a}}+
c_{3} e^{\frac{1}{2} t (-2-\nu b)}
\end{eqnarray}
where 
\begin{eqnarray}
c_{1}&=& \frac{\mu \left(\mu+\left(a^2+\mu\right) \nu b\right)}{\left(a+a^2 \mu+\mu^2\right) (2+\nu b)},\\
c_{2}&=& -\frac{a}{\left(a+a^2 \mu+\mu^2\right) \left(-1+\frac{a (-1+a \mu)}{-\mu^2+a \nu b}\right)},\\
c_{3}&=& -\frac{\mu \left(\mu+a^2 \nu b+\mu \nu b\right)}{\left(a+a^2 \mu+\mu^2\right) (2+\nu b)}+ \\ \nonumber
&&\frac{a}{\left(a+a^2 \mu+\mu^2\right) \left(-1+\frac{a (-1+a \mu)}{-\mu^2+a
\nu b}\right)}.
\end{eqnarray}
Here  $\sum_{i=1}^3 p_{i}(t)=1$ revealing the probability distribution is normalized.  In the limit of $t \to \infty$, we recapture the steady state probability distributions
\begin{eqnarray}
p_{1}^{s}&=&\frac{a}{a+a^2 \mu +\mu ^2},\\
p_{2}^{s}&=&\frac{\mu  \left(2 a^2+\mu \right)}{\left(a+a^2 \mu +\mu^2\right) (2+b \nu )},\\
p_{3}^{s}&=&\frac{\mu  \left(\mu +b \left(a^2+\mu \right) \nu \right)}{\left(a+a^2 \mu +\mu ^2\right) (2+b \nu)}.
\end{eqnarray}

The velocity  $V(t)$ at any time $t$ is the difference between the forward $V_{i}^{+}(t)$ and backward $V_{i}^{-}(t)$ velocities at each site $i$ 
\begin{eqnarray}
V(t)&=& \sum_{i=1}^{3}(V_{i}^{+}(t)-V_{i}^{-}(t)) \\ \nonumber
&=&(p_{1}P_{21}-p_{2}P_{12})+(p_{2}P_{32}-p_{3}P_{23})+\\ \nonumber
&&(p_{3}P_{13}-p_{1}P_{31}).
\end{eqnarray}
Exploiting Eq. (63), one can see that the particle attains a unidirectional current when  $f=0$ and  $T_{h}>T_{c}$.  For isothermal case $T_{h}=T_{c}$, 
   the system sustains a non-zero velocity  in the presence of load $f \ne 0$ as expected. Moreover, when  $t \to \infty$,  
the velocity  $V(t)$ increases with $t$ and approaches the steady state velocity 
\begin{eqnarray}
V^{s}=3{\mu \left(b a \nu-{\mu\over a}\right) \over 2(2+\nu b)\left(1+a\mu+{\mu^2\over a}\right)}.
 \end{eqnarray}

\section*{ Appendix A2}
The expressions for $p_{1}(t)$, $p_{2}(t)$  and $p_{3}(t)$ as well as $V(t)$  are derived considering a Brownian particle that operates in a heat bath where its temperature decreases linearly along with the reaction coordinate.
For the particle which is initially  situated at  site $i=1$,  the time-dependent  normalized probability distributions after solving  the rate equation 
  ${d \vec{p}  \over dt}= {\bold P}\vec{p}$ are given as 
\begin{eqnarray}
p_1&=&{a_2 (2+\nu) \over \mu_2^2+a_1 a_2 \mu_1 \nu+\mu_2^2 \nu}c_1+ \\ \nonumber
&&\left(-1+{-a_2+a_1 a_2\mu_1 \over -\mu_2^2+a_2\nu}\right)  e^{\left[t\left(\frac{-a_2-a_1 a_2 \mu_1-\mu_2^2}{2 a_2}\right)\right]}c_2 \\
p_2&=&-\frac{-2 a_1 a_2 \mu_1-\mu_2^2}{\mu_2^2+a_1 a_2 \mu_1 \nu+\mu_2^2 \nu}c_1- \\ \nonumber
&&\frac{-a_2+a_1 a_2 \mu_1}{-\mu_2^2+a_2
\nu}e^{\left[t\left(\frac{-a_2-a_1 a_2 \mu_1-\mu_2^2}{2 a_2}\right)\right]}c_2-\\ \nonumber
&&e^{\left[t\frac{1}{2} (-2-\nu)\right]}c_3 \\
p_3&=&c_1+e^{\left[t\left(\frac{-a_2-a_1 a_2 \mu_1-\mu_2^2}{2 a_2}\right)\right]}c_2+e^{\left[\frac{1}{2}(-2-\nu)t\right]}c_3
\end{eqnarray}
where 
\begin{eqnarray}
c_1&=& -{-\mu_2^2-a_1 a_2 \mu_1 \nu-\mu_2^2 \nu \over \left(a_2+a_1 a_2 \mu_1+\mu_2^2\right) (2+\nu)} \\
c_2&=&-{a_2 \left(-\mu_2^2+a_2 \nu\right)\over \left(a_2+a_1 a_2 \mu_1+\mu_2^2\right) \left(-a_2+a_1 a_2\mu_1+\mu_2^2-a_2 \nu\right)} \\
c_3&=&{\mu_2^2-2 a_2 \nu+a_1 a_2 \mu_1 \nu+\mu_2^2 \nu-a_2 \nu^2\over (2+\nu) \left(a_1-a_1 a_2 \mu_1-\mu_2^2+a_2 \nu\right)}.
\end{eqnarray}
Once again,   $\sum_{i=1}^3 p_{i}(t)=1$ revealing the probability distribution is normalized. 
When $t \to \infty$, the steady state probability distributions converge to  
\begin{eqnarray}
p_{1}^{s}&=&{a_2\over (a_2 + a_1 a_2 \mu_1 + \mu_2^2)},\\
p_{2}^{s}&=&{(2 a_1 a_2 \mu_1 + \mu_2^2)\over ((a_2 + a_1 a_2 \mu_1 + \mu_2^2) (2 + \nu))},\\
p_{3}^{s}&=&{(a_1 a_2 \mu_1 \nu + \mu_2^2 (1 + \nu))\over ((a_2 + a_1 a_2 \mu_1 + \mu_2^2) (2 + \nu))}.
\end{eqnarray}
The velocity  $V(t)$ at any time $t$ is the difference between the forward $V_{i}^{+}(t)$ and backward $V_{i}^{-}(t)$ velocities at each site $i$ 
\begin{eqnarray}
V(t)&=& \sum_{i=1}^{3}(V_{i}^{+}(t)-V_{i}^{-}(t)) \\ \nonumber
&=&(p_{1}P_{21}-p_{2}P_{12})+(p_{2}P_{32}-p_{3}P_{23})+\\ \nonumber
&&(p_{3}P_{13}-p_{1}P_{31}).
\end{eqnarray}
In the limit  $t \to \infty$,  
the velocity  $V(t)$ increases with $t$ and approaches to steady state velocity
\begin{eqnarray}
V^s={(3 (-\mu_2^2 + a_1 a_2 \mu_1 \nu))\over (2 (a_2 + a_1 a_2 \mu_1 + \mu_2^2) (2 + \nu))}.
\end{eqnarray}

\section*{Acknowledgment}
I would like to thank Blaynesh Bezabih and Mulu  Zebene for their
constant encouragement.

\end{document}